\pdfoutput=1
\documentclass{article}

\usepackage{graphicx}
\usepackage{algorithm}
\usepackage{algorithmic}
\usepackage{amsfonts}
\usepackage{graphicx}
\usepackage{float}
\usepackage{booktabs}
\usepackage{setspace}
\usepackage{changes}
\usepackage{amsmath,amsthm}

\begin{document}

\title{A Weight-based Information Filtration Algorithm for Stock-Correlation Networks}
\author{Xue Guo$^{1,2}$, Hu Zhang$^2$ and Tianhai Tian$^{(3,*)}$\\
$^1$School of Economics, Wuhan Textile University, China\\
$^2$School of Statistics and Mathematics, \\
Zhongnan University of Economics and Law, China\\
$^3$School of Mathematics, Monash University, Australia,      \\
 $^*$Corresponding author\\
 \{snowygx@126.com, zhh11497@sina.com, tianhai.tian@monash.edu\}}
\date{}
\maketitle

\begin{abstract}
Development of stock networks is an important approach  to explore the relationship between different stocks in the era of big-data. Although a number of methods have been designed to construct the stock correlation networks, it is still a challenge to balance the selection of prominent correlations and connectivity of networks. To address this issue, we propose a new approach to select essential edges in stock networks and also maintain the connectivity of established networks. This approach uses different threshold values for choosing the edges connecting to a particular stock, rather than employing a single threshold value in the existing asset-value method. The innovation of our algorithm includes the multiple distributions in a maximum likelihood estimator for selecting the threshold value rather than the single distribution estimator in the existing methods. Using the Chinese Shanghai security market data of  151 stocks, we develop a stock relationship network and analyze the topological properties of the developed network. Our results suggest that the proposed method is able to develop networks that maintain appropriate connectivities in the type of assets threshold methods.

 \vspace{3mm} {\bf Key Words:} Mutual Information, Threshold, Maximum likelihood estimation, Clique
\end{abstract}

\section{Introduction}
Complex system consists of a large number of components that interact with each other. 
It is important to identify the influence of each node on the dynamics of other nodes by using the relationship between different nodes. A wide variety of applications have been conducted for developing various network models such as social networks\cite{NEW}, biological networks\cite{ZOU},  financial networks\cite{MAN} and technological networks\cite{TUM,ALB,BAR}.\\
Financial markets have been studied as financial networks with fluctuationg interdependencies of the asset pricing\cite{SPS}. A typical case is the stock market, in which stocks affect each other according to the national policies, industrial development, business performance and occasional events. The correlation-based network has become an effective way to study the structure of stock markets\cite{MAN,ONN, ONN2, CHI}. Some common characteristics of stock networks have been found, such as small world\cite{CLA,WDJ, BVB, AJB} and scale free\cite{CHI,CGS}. According to the comparison of topological properties in different periods, the efficiency and instability have been growing in the stock market\cite{CTM}. It has different structures around the financial crisis\cite{HXX} and takes on more concentrated topological structure in financial crisis than in other time periods\cite{HRH, WXS}. In addition, a stock network may be fragile to targeted attacks and meanwhile may have topological robustness\cite{WLZ, HWQ}. These topological analysis results are considerably useful in portfolio optimizations\cite{MAN, ZWW}. \\
The initial associated network constructed by the correlations between stock prices is a complete network. The common objective of correlation networks is to extract a representative subgraph with essential information from the whole associated network. Currently, there are three major methods to find the crucial information to form a sub-graph, namely the minimum spanning tree (MST) \cite{ATM}, planar maximally filtered graph (PMFG) \cite{TUM,TAM}, and asset graph based on the threshold value method \cite{ONN}. MST extracts a general hierarchical structure \cite{BCL} by connecting $n$ nodes with $n-1$ edges without any loop in the network. MST probably has a severe reduction of edges in order to keep the whole weights of network as the minimum. However, the removal of a large number of edges may lead to the loss of valuable information \cite{TUM}. PMFG is a graph embedded on a surface with certain genus, which decides the complexity of graph. PMFG can supply more information associated with loops and cliques by increasing the value of genus, but some  major correlations may still be deleted from the network in order to keep the graph plane. Compared to these two methods, the threshold graph is a more acceptable method, which is easier to obtain a filtered network by adding edges whose correlations are above   a pre-selected threshold value \cite{ONN2}. The complexity of considered network can be determined by varying the  threshold value \cite{BPS}. It has been found that the majority of stocks in the market rely on a small number of close connected stocks within the same financial sector \cite{CHI} and the topology of the threshold graph is relatively stable in both of normal and crashing markets \cite{OKK}. In addition, a threshold graph presents clusters earlier and has less scale-free property than the MST. However, the threshold graph favors the most relevant correlations regardless of the structure network since  some nodes may be excluded from the network. \\
The objective of this paper is to develop an effective method for filtering pertinent information in order to observe clusters in the network in view of homogeneity among stocks. Since stocks in different sectors may have multifarious levels of relevance, some stocks may be excluded from the network if a fixed threshold value is applied to all correlations. To address this issue, we propose a new methodology which leads to an optimal structure with all stock nodes by using different threshold values for the correlations within different stocks. The maximum likelihood estimation method is used to determine the threshold values, which has been used to determine the cut-off value for selecting samples of a given distribution \cite{XGL}.  We have used this method recently to select the optimal threshold values for each stock based on the Gaussian distribution \cite{XHJ} . However, our research suggested that a single distribution was not appropriate to model samples with both smaller values and larger values. In this work we propose a maximum likelihood method with two distributions to model samples with distinct correlations. In addition, we further introduce constraint in the new method to adjust the selection of edges with close correlations. \\
The following part of this paper is presented as follows: Section 2 introduces the data set and correlation measures between stocks. Section 3 proposes two new approaches for selecting threshold values to develop stock networks, namely the likelihood threshold method and the constrained likelihood threshold method. In Section 4, we compare stock networks based on these methods, and study the topological properties of these networks. Section 5 is the conclusion of this work.\\

\section{Data set}
\subsection{Sample selection}
Shanghai Stock Exchange (SSE) in China is composed of multiple enterprises from different industries. In this study we use the dataset from the SSE 180 Index which is the stock index representing the top 180 companies by "float-adjusted" capitalization and other criteria. SSE 180 is a sub-index of SSE Composite Index, the latter included all shares of the exchange. The SSE 180 is reviewed every half year, and stocks may be added to or removed from the index based on the financial performance of the companies. Therefore, the sample used in our study includes a total of 151 stocks rather than 180 based on the completeness of the data in the time period  from 2014 to 2018, referring to 1157 observations of each stock returns. These 151 stocks are classified into 13 general categories according to Industrial Classification for national economic activities, which are Financial Industry (34 stocks), Electricity, Thermal, Gas and Water Production and Supply Industries (6 stocks), Transportation, Warehousing and Postal Services (8 stocks), Manufacturing Industry (55 stocks), Mining Industry (9 stocks), Real Estate (11 stocks), Information Transmission, Software, Information Technology Service (7 stocks), Construction Industry(8 stocks), Wholesale and Retail Trade Industry (8 stocks), Culture, Sports and Entertainment (2 stocks), Agriculture, Forestry, Animal Husbandry Industry (1 stock), Composite Industry (1 stock), Leasing and Business Services (1 stock). In order to distinguish their attributions, we label the nodes with different colors in the graph, which are Financial Industry (FI, red), Electricity, Thermal, Gas and Water Production and Supply Industries (ETGW, brown),Transportation, Warehousing and Postal Services (TWP, white), Manufacturing Industry (MA, purple), Mining Industry (MINI, gray), Real Estate (RE, black), Information Transmission, Software, Information Technologg Service (IT, blue), Construction Industry (CO, orange), Wholesale and Retail Trade Industry (WR, pink), Culture, Sports and Entertainment (CSE, mauve), Agriculture, Forestry, Animal Husbandry (AFAH, plum), Composite Industry (CI, turquoise), Leasing and Business Services (LBS, yellow).
In order to distinguish their attributions, we label the nodes with different colors in the graph, which are Financial Industry (FI, red), Electricity, Thermal, Gas and Water Production and Supply Industries (ETGW, brown),Transportation, Warehousing and Postal Services (TWP, white), Manufacturing Industry (MA, purple), Mining Industry (MINI, gray), Real Estate (RE, black), Information Transmission, Software, Information Technolog Service (IT, blue), Construction Industry (CO, orange), Wholesale and Retail Trade Industry (WR, pink), Culture, Sports and Entertainment (CSE, mauve), Agriculture, Forestry, Animal Husbandry (AFAH, plum), Composite Industry (CI, turquoise), Leasing and Business Services (LBS, yellow). \\
\subsection{Measure of correlations between stocks}
To compose a stock correlation network, we start with the mutual dependency between each stock pair in a stock portfolio, which has been universally quantified by the correlation coefficient\cite{MAN, ONN, CHI} and partial  correlations\cite{KTM, CTM} . This measure mostly describes linear relationships and does not satisfy the demand for practical problems. For example, the Chinese stock market had experienced sharp fluctuations from 2014 to 2017.  During that time period, most stock prices multiplied and went down to the original price afterwards, leading to notable nonlinear trends between stock pairs. Therefore, we explore mutual information (MI) to measure the nonlinear relationship between stocks, based on Entropy Theory\cite{JMK}.  MI has been widely applied to biological data analysis, which can explain different kinds of relationships, such as exponential, quadratic curve and linear relations. It has also been applied to quantify the correlations between stocks \cite{XG}. The MI of two stocks is estimated as follows. The logarithm return would be applied instead of stock price. The logarithm return of  stock  $i$ on day $t$ is defined as
\begin{eqnarray}
S_{i,t} &=& ln\frac{p_{i,t}}{p_{i,t-1}}, (t=2,...T; i=1,2...,n),
\end{eqnarray}
where $p_{i,t}$ is the closing price of stock $i$ on day $t$.   
\paragraph{}For a discrete variable $X$, the entropy $H(X)$ is
\begin{equation}\label{eq1}
H(X)=-\sum_{x\in X} p(x)\log p(x).
\end{equation}
where $p(x)$ is the probability of each discrete value $x$ in $X$. The joint entropy $H(X,Y)$ of random variables $X$ and $Y$ can be denoted by
\begin{equation}\label{eq2}
H(X,Y)=-\sum_{x\in X, y\in Y} p(x,y)\log p(x,y).
\end{equation}
where $p(x,y)$ is the joint probability of $x$ in $X$ and $y$ in $Y$. Based on these definitions, the
 mutual information between stock $i$ and $j$ can be estimated by
\begin{eqnarray}
I(S_{i},S_{j}) &=& H(S_{i})+H(S_{j})- H(S_{i},S_{j}), (i,j=1,2...,n).
\end{eqnarray}
Here,  $H(S_{i})$ is the entropy of stock $i$ and $H(S_{i},S_{j})$ is the joint entropy of stocks $i$ and $j$. $I(S_{i},S_{j})$ means the common information that stocks $i$ and $j$ share. The result of $I(S_{i},S_{j})$ takes a value in $[0,+\infty)$ and a larger value corresponds to a closer relationship. Usually the normalized MI is more commonly used, which is defined as 
\begin{eqnarray}
MI(S_{i}, S_{j}) &=& \frac{I(S_{i},S_{j})}{H(S_{i},S_{j})}, (i,j=1,2...,n).
\end{eqnarray}
where  $MI \in [0, 1]$.
In developing a network, the distance of two stocks is transformed by 
\begin{eqnarray}
D(S_{i}, S_{j}) &=& 1-\frac{I(S_{i},S_{j})}{H(S_{i},S_{j})}, (i,j=1,2...,n).
\end{eqnarray}
Formula (4) indicates that shorter distances correspond to stronger correlations. For each pair of stocks, we can get their MI and distance correspondingly. Therefore, the symmetric matrices of mutual information $MI_{n\times n}$ and distances $D_{n\times n}$ can be explored by formulas (3) and (4), respectively.

\section{Methodology}
\subsection{Traditional threshold method}
The basic idea of traditional threshold method is to select the strongest links with the largest values of correlations to form a network.  According to formula (4), the distance matrix $D_{n\times n}$ is used to determine topological structure connecting $n$ stocks in a certain portfolio. In the previous research \cite{ONN,CK}, all values in matrix $D_{n\times n}$ are sorted in an ascending order $\{d_{(1)},d_{(2)},\cdots,d_{({n\times(n-1)/2})}\}$. Given a threshold $d^*$, these values are divided into two parts, and the distances which are less than $d^*$ will be included in the threshold graph. Correspondingly, a selected set $E$ consists of links whose values are above a certain value and stock pairs in set $E$ have stronger relationship than the other stock pairs. This Algorithm\ref{tab:table1} is described as follows.
\begin{table} [h]
	\centering
      \caption{\label{tab:table1}Algorithm 1}
     \begin{tabular} {l}
		\hline
		Threshold algorithm \\ \hline
Input: normalize mutual information matrix $MI_{n\times n}$ \\
 \quad (or distance matrix $D_{n\times n}$), and the node set $V$\\
Output: Edge set $E$ connecting nodes in $V$\\
Step 1: sort values in $MI_{n\times n}$ in a descending order \\
\quad (or $D_{n\times n}$ in an ascending order)\\
Step 2: Set a threshold $\eta^*$ for $MI_{n\times n}$ (or $d^*$ for $D_{n\times n}$)\\
Step 3: for $i=1:n$ \\
\quad for $j=i+1:n$ \\
\qquad  if  $MI(i,j)> \eta^*$(or $d(i,j)<d^*$)\\
  \qquad  \quad          Add $e(i,j)$ to set $E$\\
\qquad            endif\\
 \quad      endfor\\
endfor\\
Step 4: Use $E$ to plot the graph of the established network. \\
		\hline
	\end{tabular}	
\end{table}
As mentioned above, the traditional threshold method focuses on the strong relationship and intensive clusters among stocks. As a result, a proportion of links will be removed because of the small values of correlations, though some of them are also important to the network. For some stocks in the Transportation, Wharehousing and Postal Services Sector, for example, their prices are quite stable in any time period, even in a cycle of economic boom or in financial crisis. This 
results in a lower overall relevance of stock pairs between this sector and other sectors. Thus, the stock nodes will be excluded from the network when the threshold value gets larger. However, 
if a relatively smaller value of threshold is chosen to include these stocks, the network will be dense and it would be difficult to derive major information from the network. 
\subsection{Likelihood method using multiple distributions}
Based on the discussion in previous subsection, a measure should be applied not only to solve the problem of excluded nodes but also to keep the strong correlations in the graph. Thus, stocks in different sectors having distinctive levels of correlations should have  varied levels of thresholds  in order to classify correlation values into strong part and weak part.  For each stock, we will set up a corresponding threshold value.  \\
Firstly, we sort the $MI_{i,j}$ values of the stock $i$ with all other stocks in an ascending order. Then a threshold value should  be determined for each stock node rather than a unified threshold for all nodes. For stock node $i$, vector $X_{i}=(x_{i,1},\cdots,x_{i,n-1})$ represents the MI values in an ascending order. Then we use a breakpoint $u$ to divide the vector into two parts, the weak correlation part $E_{weak} = \{x_{i,1},\cdots,x_{i,u}\}$ and strong correlation part $E_{strong} = \{x_{i,u+1},\cdots, x_{i,n-1}\}$. Nodes related to $E_{strong}$ should be added to the target node set $V$ and links in  $E_{weak}$  should be filtered out. Then the issue is how to set up the point $u$ to distinguish them. 
\paragraph{}To address this issue, a method using the Maximum Likelihood Estimate (MLE) has been proposed to use a single distribution to classify these values \cite{XGL, XHJ}. However, our research results suggest that this single distribution is not accurate to calculate the likelihood related to the strong correlation part \cite{XHJ}. Here we propose to use  two distributions with different characteristics to provide a more accurate classification. The best division should be inclined to make two distributions having the biggest difference of MLE values.
Using the notation above,  the maximum likelihood function is defined as
\begin{eqnarray}
ML(u)=&&\mbox{log}(L_1((x_{i,1},\cdots, x_{i,u})|\theta_1))+\\
&& \mbox{log}(L_2((x_{i,u+1},\cdots, x_{i,n-1})|\theta_2)), \nonumber
\end{eqnarray}
where $L_1$ and $L_2$ are two different likelihood functions with distinct parameters $\theta_1$ and $\theta_2$ with respect to $E_{weak}$ and $E_{strong}$, respectively.

Now the main problem is the choice of these distributions. The normal distribution is a common approach if the amount of data is comparably large, but it may not be accurate when the amount of data is quite small. Here, we simulate $X_{i,1:u}$ and $X_{i,u+1:n-1}$ independently by frequency distribution fittings and test their siginificance of distributions, such as normal, possion, exponential and rayleigh distributions. Thus it is called the Multi-Likelihood Method (MLM), which is given in Algorithm\ref{tab:table2} .

\begin{table} [h]
	\centering
      \caption{\label{tab:table2}Algorithm 2}
     \begin{tabular} {l}
		\hline
		Multi-Likelihood Method (MLM) \\ \hline
Input: normalize mutual information matrix Matrix $MI_{n\times n}$\\
 \quad and the node set $V$\\
Output: Edge set $E$ connecting nodes in $V$\\
for $i=1:n$ \\
\quad Sort the values of MI in the $i$-row to get vector $X_i$ \\
\quad Find the optimal breakpoint $u_{i}$ using (5).\\
\quad for $j=i+1:n$\\
\qquad  if  $MI(i,j)> u_{i}$\\
  \qquad    \quad       Add $(i,j)\in V$ and $e(i,j)\in E$\\
\qquad            endif\\
 \quad      endfor\\
endfor\\
Use $E$ to plot the graph of the established network. \\
		\hline
	\end{tabular}	
\end{table}
\subsection{Constrained Multi-Likelihood Method}
Although the proposed Algorithm\ref{tab:table2} is able to solve the problem of excluded nodes, our tests suggest that the derived networks may contain as relevant  information as possible. To derive a network with appropriate number of links for each stock, a penalty function $g(x_{i})$ is embedded into the likelihood function, which is composed by constraining the total weights of selected links. This consideration leads to the following Constrained Multi Likelihood Method (CMLM)
\begin{eqnarray}
CML(u)=&&\mbox{log}(L_1((x_{i,1},\cdots, x_{i,u})|\theta_1))\nonumber\\
&&+ \mbox{log}(L_2((x_{i,u+1},\cdots, x_{i,n-1})|\theta_2))-\alpha \times g(x_{i}), 
\end{eqnarray}
Here $L_1$ and $L_2$ are different likelihood functions for weak links and strong links, respectively, $\alpha$  is  a regularized parameter which can adjust the number of links included in the network. When $\alpha$ increases, some edges related to small values of MI will be gradually removed from the network. The Algorithm\ref{tab:table3} is given  below.

\begin{table} [h]
	\centering
      \caption{\label{tab:table3}Algorithm 3}
     \begin{tabular} {l}
		\hline
		Constraint Multi-Likelihood Method (CMLM) \\ \hline
Input: normalize mutual information matrix Matrix $MI_{n\times n}$\\
 \quad and the node set $V$\\
Output: Edge set $E$ connecting nodes in $V$\\
for $i=1:n$ \\
\quad Sort the values of MI in the $i$-row to get vector $X_i$ \\
\quad Calculate $CML$ using formula (6). \\
\quad Find the optimal breakpoint $u_{i}$ using $u=\mbox{argmax}(CML)$.\\
\quad for $j=i+1:n$\\
\qquad  if  $MI(i,j)> u_{i}$\\
  \qquad\quad       Add $(i,j)\in V$ and $e(i,j)\in E$\\
\qquad            endif\\
 \quad      endfor\\
endfor\\
Use $E$ to plot the graph of the established network. \\
		\hline
	\end{tabular}	
\end{table}
The key question in the Algorithm \ref{tab:table3} is the selection of the regularized parameter $\alpha$ and function $g(x_{i,j})$, which will be discussed in detail in the following section. \\
\section{Results and Discussions} 
\subsection{Distributions of MI values}
Based on our sample data, there are totally 11325 (namely $C_{151}^{2}$) values of MI for all the stock pairs, ranging from 0.0308 to 0.7092 with the average 0.1584 and the median 0.1520. The ranges and average values of MI for the 10 major sectors are given in Table \ref{tab:table4}. The distributions of  the MI values is uneven. Among them, $84.26\%$ of correlations take values from 0.1 to 0.3 while only $2.02\%$ of them are over 0.3.  Table \ref{tab:table4} also shows that the FI, IT and CO sectors have higher lever of average correlations than the other sectors while the WR and CSE sectors have lower levers. Meanwhile, the FI, TWP, MIN and CO sectors have larger deviations of MI value than other sectors.
\begin{table} [h]
	\centering
      \caption{\label{tab:table4}Distributions of  the MI values for the 10 major sectors}
     \begin{tabular} {lcc||lcc}
		\hline
		Sector  & Range of MI &Average MI& Sector  &Range of MI&Average MI  \\ \hline
		FI  &$[0.0308, 0.6546]$&0.1648&TWP&$[0.0523, 0.7092]$&0.1554 \\
		MA &$[0.0356, 0.5365]$ &0.1547&MIN&$[0.0378, 0.6346]$ &0.1569\\
		RE  &$[0.0336, 0.4110]$&0.1557&IT &$[0.0472, 0.3176]$&0.1606\\
		CO  &$[0.0566, 0.6399]$&0.1840&WR &$[0.0459, 0.3367]$&0.1476\\
             ETGW&$[0.0308, 0.3367]$&0.1539&CSE&$[0.0462, 0.2507]$&0.1224\\
		\hline
	\end{tabular}	
\end{table}
\subsection{Networks using the threshold algorithm}
\begin{figure}
 	\centering
 	\includegraphics[width=9cm]{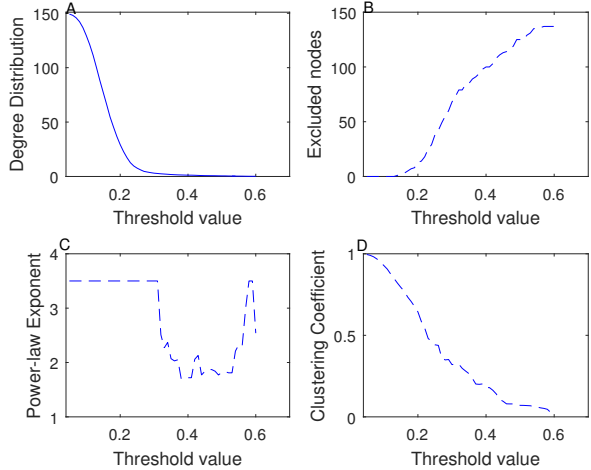}
 	\caption{\label{fig:epsart}Topological properties of the stock networks derived from Algorithm 1. (A-D)  show the average degree, number of excluded nodes, power-law exponent $\gamma$, and clustering coefficient for networks determined by different threshold values, respectively.}
 \end{figure}

Following Algorithm \ref{tab:table1}, we first construct a network by giving a threshold with value $\eta$. For $\eta \in (0.05, 0.6)$, the number of edges decreases as $\eta$ increases. The structure of network is not well defined if the value of $\eta$ is too small or too large. Figure \ref{fig:epsart} demonstrates the variations of network topology with $\eta$ increasing. 
\\
When $\eta \in (0.05, 0.20)$, the degree distribution is approximately a straight line and decreases slowly afterwards because most of correlations gather at  threshold interval (0.05, 0.20). However, some nodes is excluded from the network if $\eta$ is over 0.14. For $\eta \in (0, 0.14)$, all the nodes are included in the network but the network has a relatively large value of degree which is over 100. \\ 
In accordance with Vandewalle's discovery \cite{VBT}, many real-world networks are scale-free, which means that only a few nodes should have more links while the others have relatively few links. The power-law function can appropriately describe the degree distribution of a real network, given by $$p(k)\sim k^{-\gamma}$$ where $k$ is the value of degree, and $p(k)$ represents the proportion of the k-degree nodes. Usually, the network is called scale-free if $\gamma\in (2, 3)$, which reflects that the notable characteristic of most nodes have uniform degree distribution and only few nodes have large degree.  As shown in Figure \ref{fig:epsart}C, the network is scale-free only when $\eta \in (0.32, 0.57)$.
\\
Clustering, originating from the percolation theory\cite{HJJ}, is a convincing characteristic in stock networks that some units closely connect to each other. A cluster means a group of three stock nodes that connect each other, forming a strong unit. The clustering coefficient is applied to describe the clustering level of the graph, which is defined as the ratio of the number of existing triangles to the number of all possible triangles. The clustering coefficient of networks in Figure 1D is getting smaller with the increase of threshold values. In particular, it drops sharply when the value of $\eta$ ranging from 0.05 to 0.2. Compared to the cases with $\eta \in (0.40, 0.60)$, the clustering coefficient of networks with $\eta \in (0.05, 0.20)$ is much larger. Thus, it is clear that the topology of networks is highly sensitive to the value of threshold $\eta \in (0.05, 0.20)$.  However, the network is not completed with $\eta \in (0.14, 0.60)$ since some nodes are disconnected from the network. As a result, it is difficult to select a proper threshold value in the traditional threshold value framework in order to generate a network with both good edge density and completeness of the network. 
\subsection{Network using Multi-Likelihood Method}
\begin{figure}[t]
 	\centering
  	\includegraphics[width=9cm]{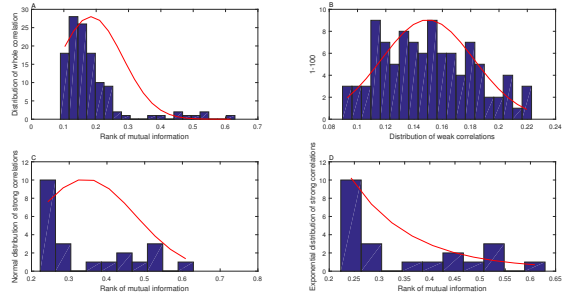}
  	\caption{\label{fig:dist} Different distributions to approximate the data frequency of stock "Sany Industry". A. Frequency of the total MI values between stock Sany and other stocks. B. Distributions for weak correlation. C. Use normal distribution to fit the strong correlation. D. Use the exponential distribution to fit strong correlations.}
  \end{figure}

We have shown in Table \ref{tab:table4} that different sectors have different average values of MI. Thus, it is not appropriate to apply a single  threshold to all sectors and to all nodes. A natural idea is to set a threshold value for a sector or for a node individually. 
A series of threshold values can be detected following Algorithm 2. Strong correlations could be distinguished by formula (7). Usually, $L_{1}(x\mid\theta_{1})$ and $L_{2}(x\mid\theta_{2})$ are supposed to be based on the normal distributions \cite{XGL}. However, the normal distribution may not be able to fit every sample dataset. In financial areas, the distribution of logarithmic returns shows the characteristic of a peak and long tail because of extreme values. Thus we need to  find other distributions to approximate the distribution of correlations more accurately. We apply several types of distributions to test the frequency of correlations, such as the normal, Poission, exponential and Rayleigh distributions. The results, for strong correlations, show that the exponential distribution fits the data with the highest accuracy. As an example, Figure  \ref{fig:dist} demonstrates a comparison of distributions for stock "Sany Industry". It is evident that the exponential distribution in Figure \ref{fig:dist}D  fit the samples better than the normal distribution in Figure  \ref{fig:dist}C.\\
Then we need to find out the breakpoint for each stock. According to formula (5), most thresholds take values in the interval $(0.1, 0.2)$ and only a few thresholds are less than 0.1, resulting in 9981 links included in the graph. Based on MLM, strong correlations are gradually selected for each stock. The network is more homogeneous compared to the graphs constructed by the traditional threshold method. It should be noted that there are a large number of edges in the network due to the small threshold values.\\

\subsection{Network using Constrained Multi-Likelihood Method}
To reduce the number of edges in the networks in previous subsection, a method should be designed in order to get an optimised network which can connect all nodes and has a good distribution of degrees. According to Algorithm 3, we consider a penalty function  $\alpha \times g(x_{i})$  as a constraint factor embedded into the likelihood function in order to filter out further information. In this work we consider the following function 
$$\alpha \times g(x_{i})=\alpha\frac{\sum_{j=u+1}^{n-1}(1-x_{i,j})}{(\frac{1}{n-1}\sum_{j=1}^{n-1}x_{i,j})^q},$$
where $(0\leq\alpha<1,q\geq1).$ As the values of $\alpha$ and $q$ increase, less links will be included in the graph. When $\alpha$ equals to 0, this measure is equal to that in MLM. We have tested different values of $q$ and find that the network has appropriate distribution of degrees when the value of $q$ is set to 2.

\begin{figure}
	\centering
	\includegraphics[width=9cm]{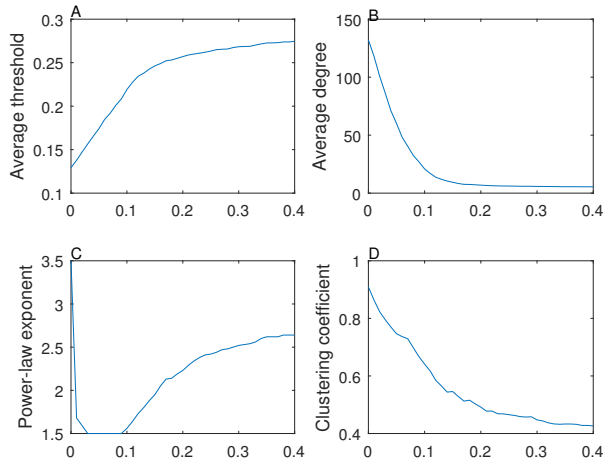}
	\caption{\label{fig:CMLM} Topological properties of the stock networks derived from CMLM. (A-D)  show the average degree, number of excluded nodes, power-law exponent $\gamma$, and clustering coefficient for networks determined by different values of $\alpha$, respectively. }
\end{figure}


Figure \ref{fig:CMLM} provides the topological properties of the derived networks with $\alpha$ increasing from 0 to 0.4.  The average value of threshold increases from 0.1293 to 0.2745 when $\alpha$ increases from 0 to 0.4. As the number of edges is reduced in the graph, the average degree goes down dramatically. While $\alpha$ increases by 0.01, the links of each stock averagely drop by 15. On average, the links of each node decline from 132.1987 to 5.4702. Thus the network structure changes with the variation of $\alpha$, leading to a wide range of power-law exponent within $[1.5, 3.5]$. The network is scale-free when $\alpha$ takes a value in the interval $[0.22, 0.4]$. The power-law property becomes more evident when the number of edges is getting less. In addition, the clustering coefficient has the similar tendency with that of the average degree. However, the clustering coefficient has less variations, dropping from 0.9092 to 0.4263. This method is capable of achieving a simpler topology containing the most relevant edges for each stock node. Figure 4 gives the stock network using the CMLM method based on $\alpha=0.3$. Note that this figure cannot be published on arXiv, but it will be published later.

\subsection{Properties of cliques}
A clique $K_m$ is a subset of $m$ nodes in which each node directly connects the other nodes within the subset \cite{PDF, TN}. Stocks in the same clique would have stronger mutual influences than the stocks outside of this clique. We then study cliques in the network developed by using the CMLM method (namely Figure 4). There are total 437 links in the graph and 77 cliques, ranging from 3 to 10 elements. $K_m$ $(m\geq5)$ account for $1/5$ of total cliques while the others are 3-cliques and 4-cliques. \\
We first study cliques in terms of classifications of sectors. The analysis on cliques reveals a highly homogeneous trend with respect to industrial sectors. According to statistics, 34 cliques out of the 77 cliques contain stocks belonging to the same sector, 34 cliques are composed of stocks from 2 sectors, but only 9 cliques have stocks from 3 sectors. Table \ref{tab:table5} lists the information of large cliques ($m\geq 5$). The largest 4 cliques ($m\geq 8$) include stocks belonging to Financial sector, and one of 7-cliques is composed by stocks in Construction sector.  \\
To study the topology of cliques, we next consider the statistical property named disparity \cite{TAM}, which is a quantity as the average value of the disparity measure inside a clique, defined by
\begin{equation}
y(i)=\sum_{j\neq i, j\in clique}(\frac{MI_{ij}}{s_i})^{2},
\end{equation}
where $s_i =\sum_{j\neq i, j\in clique}(MI_{ij}) $.
The network is detected to be hierarchical since cliques have varied ranges of similarity and disparity. In particular, the financial sector and construction sector have stronger correlations. The maximum average correlation is 0.5289 showing in the 10-clique while the minimum average correlation is 0.3704 in a 6-clique. In addition, the cliques have small diversities. The values of disparity range within $[0.0225, 0.1047]$. The larger clique yields the smaller disparity. 
\begin{table*}
\centering
\caption{\label{tab:table5}%
Information of cliques $K_m(m\geq5)$}
\begin{tabular}{lcccc}  
\hline 
   K-clique & Number &Sector (Frequency)& Average MI & Disparity of clique \\
  \hline
 10-clique &  1& FI(10) & 0.5289&0.0225  \\
9-clique&  2 &  FI(18) & [0.4843, 0.4991]&0.0290 \\
 8-clique &  1 & FI(8) &0.4841&0.0373  \\
 7-clique & 3& FI(6), RE(1), MA(1), CO(13) & [0.3844, 0.4977]&[0.0486, 0.0510]  \\
 6-clique & 3& FI(12), MA(1), CO(5)&[0.3704, 0.4735]& [0.0680, 0.0707] \\
5-clique&5&FI(15), MA(4), MINI(2), CO(4)&[0.3868, 0.4363]&[0.1010, 0.1047]\\
  \hline
  \end{tabular}
\end{table*}
For cliques from diverse sectors, Table \ref{tab:table6} shows that only seven 3-cliques belong to three distinctive sectors. The mean correlation of these cliques  demonstrates a large variation of taking values in $[0.2130, 0.6631]$, whereas their disparities are close to $1/3$. The majorities of inter-cliques are clustered by stocks from two sectors, such as manufacture, mining, real estate, wholesale and retail trade, and information technology sectors. 
\begin{table*}
\centering
\caption{\label{tab:table6}%
Intrasector cliques of $K_m(m=3)$}
\begin{tabular}{p{4.5cm}p{4.5cm}p{4.5cm}} 
\hline 
  Intersector & Average MI & Disparity \\
  \hline
 FI, WR, LBS   & 0.3735&0.3375  \\
ETGW, RE, WR& 0.2897&0.3342 \\
ETGW, MA, IT   &0.6631&0.3357  \\
 MA, IT, LBS & 0.2576&0.3396  \\
MA, RE, CSE&0.2130& 0.3420 \\
MA, RE, IT &0.3162&0.3484\\
RE, WR, AFAH&0.2682&0.3351\\
  \hline
  \end{tabular}
\end{table*}
Tables \ref{tab:table5} and \ref{tab:table6} illustrate that CMLM is able to select cliques at varied levels of correlation. During the investigation period, the Chinese market showed  strong homogeneous clustering. Stocks from Financial, Construction sectors are more involved in larger cliques.  In contrast, stocks from Manufacture, Mining, Real Estate, Wholesale and Retail Trade, Information Technology sectors are likely to form small cliques. Financial sector has strong levels of intrasector connections. Manufacture sector makes more interactions with other sectors.\\
Combined with the study of Tables \ref{tab:table5} and \ref{tab:table6}, we can also get main features of the cliques. Firstly, larger cliques are proved to be considerable homogeneity as they have strong correlations but small disparities. Secondly, intersector connections are mostly seen in small cliques, only 3-cliques have nodes all belonging to different sectors with the certain number of links. These features highlight the status of different sectors in the market, FI sector has strong correlations within the sector but slightly affects other sectors, MA, IT, WR and RE have more interactions cross sectors. Cliques can fully embody the interactions of distinct industries in a stock portfolio.\\
\section{Conclusion}
In this work we have studied three methods for developing stock networks based on threshold and made comparison studies of the network  structures. Our target is to construct a network containing all the nodes with clear topology properties. Using the sample data  from the SSE 180 index, we  develop networks based on the traditional threshold, MLM and CMLM methods. A number of studies have been conducted based on the traditional threshold method, which favors strong links between stocks but also excludes nodes because of the large value of the threshold. To address this issue, we have considered  networks by providing a series of threshold values for each stock node. In this way we can keep strong links with all nodes in the graph. In order to get a simplified network, a penalty function has been added to the likelihood function as a regulator. In that case, more information has been filtered out during the process of regulation. In addition, it is a good balance between links and stock nodes.  In conclusion, CMLM is an effective method to extract valuable information and include all stock nodes. The future work may be focused on the selection of the penalty function to get better topological properties of stock networks.

\end{document}